\documentclass[epj,twocolumn]{webofc}
\usepackage[varg]{txfonts}   
\usepackage{bm}

\usepackage[inline]{trackchanges}
\addeditor{LP}
\addeditor{MM}
\addeditor{RD}

\woctitle{Powders \& Grains 2017}
\begin{document}
\title{A differential equation for the flow rate during silo discharge: Beyond the Beverloo rule}
%
%

\author{\firstname{Marcos A.} \lastname{Madrid}\inst{1}\fnsep\thanks{\email{marcosamadrid@googlemail.com}} \and
        \firstname{Jos\'e R.} \lastname{Darias}\inst{2}\fnsep\thanks{\email{jrdarias@usb.ve}} \and
        \firstname{Luis A.} \lastname{Pugnaloni}\inst{1}\fnsep\thanks{\email{luis.pugnaloni@frlp.utn.edu.ar}}
}

\institute{Dpto. Ingenier\'ia Mec\'anica, Facultad Regional La Plata, Universidad Tecnol\'ogica Nacional, CONICET, Av. 60 Esq. 124, 1900 La Plata, Argentina.
\and
            Laboratorio de \'Optica y Fluidos, Universidad Sim\'on Bol\'ivar, Apartado Postal 89000, Caracas 1080-A, Venezuela. 
         }

\abstract{%
We present a differential equation for the flow rate of granular materials during the discharge of a silo. This is based in the energy balance of the variable mass system in contrast with the traditional derivations based on heuristic postulates such as the free fall arch. We show that this new equation is consistent with the well known Beverloo rule, providing an independent estimate for the universal Beverloo prefactor. We also find an analytic expression for the pressure under discharging conditions.
}
\maketitle
\section{Introduction}
\label{intro}

The discharge of grains though an opening at the base of a silo has been considered in a number of studied since the $19^{\mathrm{th}}$ century (see for example \cite{beverloo,brown1959,duran,nedderman1982,tighe2007,ristow,goldberg2016,wilson2014,mankoc2007} and references therein). The most salient feature is the fact that the flow rate does not depend on the height $h$ of the material in the container, in clear contrast with the behavior of viscous fluids. This has been sometimes attributed to the pressure saturation observed in static silos, however, discharging silos have a continuously evolving pressure \cite{aguirre2010}.

If the discharge orifice is circular and large enough to avoid clogging \cite{Zuriguel2003}, the mass flow rate $Q$ is described by the so called Beverloo rule \cite{beverloo,NeddermanBook} 
\begin{equation}
Q = C \rho_b \sqrt{g}(D_o-k\,d)^{5/2}, \label{beverloo}
\end{equation}
where $D_o$ is the diameter of the opening, $\rho_b$ the bulk density of the granular sample, $g$ the acceleration of gravity and $d$ the diameter of the grains. Here, $k$ and $C$ are two fitting dimensionless constants. Interestingly, while $k$ may vary up to a factor of 2, depending on the grains used, $C\approx 0.58$ for virtually any material tested \cite{NeddermanBook}. However, some deviations are observable for very low friction materials \cite{kondic2014}.  

The Beverloo rule is generally explained based on heuristic models such as the ``free fall arch'' model and the concept of ``empty annulus'' \cite{NeddermanBook}. However, these two models have been recently challenged \cite{janda2012,rubio2015}. Staron \textit{et al} have also shown that the Beverloo rule can be obtained if the Navier--Stokes equations are solved for a plastic fluid including a constitutive equation for the effective friction based on the $\mu(I)$-rheology \cite{staron}.

In this work, we use the global energy balance for the granular material inside a discharging silo and a constitutive relation to derive a simple differential equation for the mass $M(t)$ in the silo. We show that for a simple discharge the equation is consistent with the Beverloo rule, providing an independent estimate for the prefactor without fitting data. In the process of validating the differential equation, we also find a functional form for the pressure in the silo in the dynamic regime which is rather different to the well known Janssen equation. 

\section{Energy balance}
\label{model}

\begin{figure}
 \begin{center}
  \includegraphics[width=0.4\columnwidth]{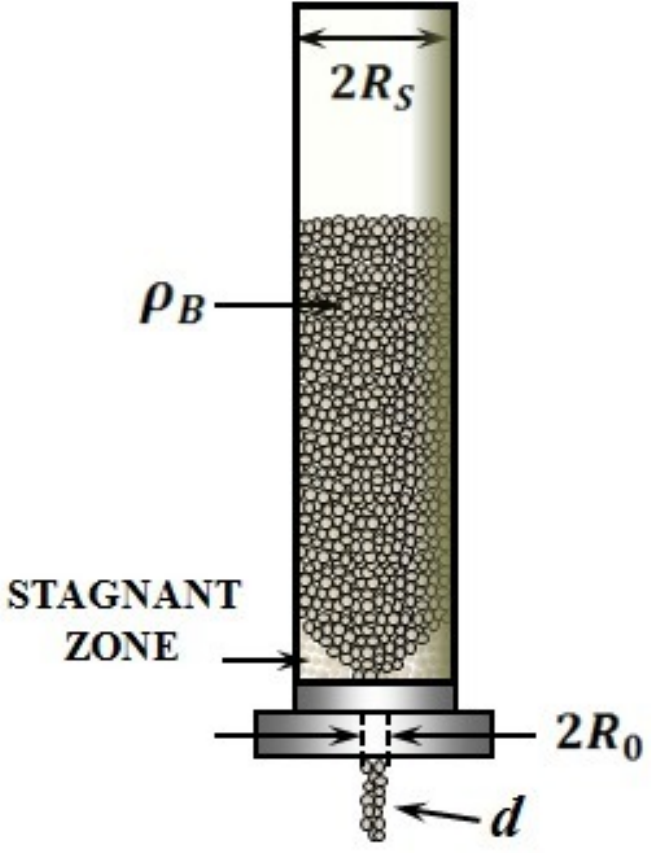}
 \end{center}
\caption{Sketch of the discharge of a silo.}
\label{set-up}
\end{figure}

Consider a cylindrical silo of radius $R_s$, discharging through an opening of radius $R_o$ (see Fig. \ref{set-up}). The energy balance requires that, at any time,
\begin{equation}
 W_g = \dot{K}_{in} + W_{out} + \dot{E} + W_D, \label{energy-balance}
\end{equation}
where $W_g$ is the power injected by the force of gravity acting on the grains, $\dot{K}_{in}$ is the rate of change of the kinetic energy of the grains inside the silo, $W_{out}$ is the power loss due to the grains that leave the silo at a velocity $v_{out}$, $\dot{E}$ is the rate of change of the elastic energy of the grains, and $W_D$ is the dissipated power due to the non-conservative interactions between grains and between grains and walls.

During the discharge, the mass $M(t)$ in the silo can be written as
\begin{equation}{\label{equiv}}
 M(t)=mN(t)\\
     =\rho_b A_s z(t)
     =2\rho_b A_s z_{cm}(t),
\end{equation}
where $m$ is the mass of one grain, $N(t)$ is the number of grains in the silo, $\rho_b$ is the apparent density in the bulk, $A_s=\pi R_s^2$ is the cross section of the silo, $z(t)$ is the head of material, and $z_{cm}(t)=z(t)/2$ is the center of mass of the granular column. We have assumed that the density is homogeneous throughout the column. Therefore, the flow rate $q(t)$ in particles per unit time is
\begin{equation}
 q(t) = \frac{Q(t)}{m}=-\frac{\dot{M}(t)}{m}=-\frac{2 \rho_b A_s}{m} v_{cm}(t), \label{q-vcm}
\end{equation}
where $v_{cm}(t)$ is the velocity of the center of mass of the granular column.

\subsection{Internal kinetic energy ($\dot{K}_{in}$)}

The kinetic energy of the grains inside the silo is
\begin{equation}
 K_{in}(t)=\frac{1}{2}\sum_{i=1}^{N(t)} m v_i^2(t),
\end{equation}
where the sum includes all particles in the silo at time $t$ and $v_i(t)$ is the velocity of particle $i$. In terms of the center of mass $K_{in}(t)$ can be expressed as 
\begin{equation} \label{kinetic-cm}
 K_{in}(t)=\frac{1}{2}M(t) v_{cm}^2(t) +  \frac{1}{2}\sum_{i=1}^{N(t)}m_i [\bm{v}_i(t)-\bm{v}_{cm}(t)]^2,
\end{equation}
where the term $[\bm{v}_i(t)-\bm{v}_{cm}(t)]^2$ can be neglected according to results obtained from DEM simulations (data not shown), hence
\begin{equation}
 K_{in}(t) \approx \frac{1}{2}M(t) v_{cm}^2(t). 
\end{equation}

Then, the rate of change of $K_{in}$ is 
\begin{equation}
\dot{K}_{in}(t)=M(t) v_{cm}(t) \dot{v}_{cm}(t) + \frac{1}{2}v_{cm}^2(t) \dot{M}(t). 
\end{equation}
which can be written, using Eqs. (\ref{equiv}) and (\ref{q-vcm}), as
\begin{equation}
 \dot{K}_{in}(t)=\frac{m^2}{4\rho^2_b A^2_s}
 \left[M(t)q(t)\dot{q}(t)-\frac{m}{2}q^3(t)\right] \label{k-in}.
\end{equation}

\subsection{Gravitational energy ($W_g$)}

The gravitational potential energy of the particles inside the silo using Eq. (\ref{equiv}) is
\begin{equation}
 U_g(t)=M(t)gz_{cm}(t)=\frac{g M^2(t)}{2\rho_b A_s}.
\end{equation}

The power injected is therefore
\begin{equation}
 W_g(t)=-\dot{U}_g(t)=\frac{m g}{\rho_b A_s} M(t) q(t). \label{wg}
\end{equation}

\subsection{Discharge energy ($W_{out}$)}

The kinetic energy that is removed from the system due to the particles that exit through the opening is
\begin{equation}
 K_{out}(t)= \frac{1}{2} m v_{out}^2 [N_0-N(t)],
\end{equation}
where $v_{out}$ is the velocity of the grains that exit the system and $N_0$ is the initial number of grains in the silo. Hence, the power removed from the system is
\begin{equation}
 W_{out}=\dot{K}_{out}(t) = \frac{1}{2} m v_{out}^2 q(t).
\end{equation}

If we consider that the mass flow rate is $mq(t)= \rho_o A_o v_{out}$, with $\rho_o$ the apparent density at the opening and $A_o$ the cross section area of the opening, then
\begin{equation}
 W_{out}=\frac{m^3 q^3(t)}{2\rho^2_o A^2_o} \label{wout}
\end{equation}

\subsection{Dissipated energy ($W_D$)}

It has been shown that for a shear cell of thickness $L$ the tangential stress $\tau$ necessary to develop a flow at velocity $v$ of a granular sample can be put in terms of the inertial number $I$ as \cite{dacruz2005}
\begin{equation}
 \tau = \mu(I)P,
\end{equation}
where $P$ is the confining pressure, $\mu(I)$ is the effective friction coefficient, $I=\frac{v}{L}\frac{d}{\sqrt{P/\rho}}$ is the inertial number that characterizes the flow if the gains are stiff and $L\gg d$, and $\rho$ is the density of the material the grains are made of.

If we assume that the flow in the silo can be represented as a shear flow, where $v$ is the velocity of the free surface, $L$ is the silo radius and $P$ is the mean hydrostatic pressure in the column (i.e., 1/3 of the trace of the stress tensor), then, the power dissipated by friction is
\begin{equation}
 W_D(t)= \tau(t) A(t) v(t) = \mu(I)P(t) A(t) v(t),
\end{equation}
where $A(t)$ is the area of frictional contact between the grains and the silo. This area includes the lateral walls, which is connected to the height of the column at time $t$, plus a fixed area that should be proportional to the base of the silo, i.e., $A(t)=2\pi R_s z(t)+\alpha\pi R_s^2 = 2\pi R_s\frac{M(t)}{\rho_b A_s}+\alpha A_s$.

Then, using Eq. (\ref{q-vcm}) and $v(t)=2v_{cm}(t)$, we obtain

\begin{equation}
W_D(t) = \mu(I)P(t) \left[2\pi R_s \frac{M(t)}{\rho_b A_s}+ \alpha A_s \right] q(t) \frac{m}{A_s\rho_b}. \label{wd}
\end{equation}

\section{Differential equation}

If the grains are stiff, the variation in the elastic energy are expected to be small and we can neglect $\dot{E}$. Notice also that $\dot{K}_{in}$ [Eq. (\ref{k-in})] falls quadratically with $A_s$ in comparison with the remaining terms of Eq. (\ref{energy-balance}) that fall linearly with $A_s$ or are independent of $A_s$. Therefore, for a wide silo $\dot{K}_{in}$ vanishes. Plugging in Eqs. (\ref{wg}), (\ref{wout}), and (\ref{wd}) into Eq. (\ref{energy-balance}) and considering Eqs. (\ref{equiv}) and (\ref{q-vcm}), we obtain

\begin{eqnarray} 
gM(t) &=& \frac{\rho_b A_s}{2\rho^2_o A^2_o}\dot{M}^2(t) \notag \\
&+& \mu(I)P(t)\left[\frac{2\pi R_s}{\rho_b A_s}M(t)+\alpha A_s\right]
\end{eqnarray}

Solving for $\dot{M}$ and using $A_o=\pi D_o^2/4$, being $D_o=2R_o$, the diameter of the opening

\begin{eqnarray}  \label{diff}
\dot{M}(t) &=& \frac{\pi\sqrt{2}}{4} \rho_o \sqrt{g} D_o^2 \left[ \frac{M(t)}{\rho_b A_s} \right.\notag \\
 &&\left. -\mu(I)\frac{P(t)}{g}\left(\frac{2\pi R_s}{\rho^2_b A^2_s}M(t)+\frac{\alpha}{\rho_b}\right) \right] ^{1/2}.
\end{eqnarray}

We note that Eq. (\ref{diff}) is similar to the Beverloo rule Eq. (\ref{beverloo}). The exponent 5/2 is not apparent since $D_o$ is squared in Eq. (\ref{diff}). We will come back to this point below. Before, it is important to mention that Eq. (\ref{diff}) includes in the prefactor $\rho_o$ instead of $\rho_b$. However, if we approximate roughly $\rho_o \approx \rho_b /2$, then the proportionality constant becomes $C = \frac{\pi\sqrt{2}}{8} \approx 0.56$. This has to be compared with the value obtained by fitting the Beverloo experiments that led to $C=0.58$, which has been shown to be remarkably insensitive to the material properties \cite{NeddermanBook}. 

Equation (\ref{diff}) is a first order differential equation for $M(t)$ that can be closed with an initial condition such as $M(t=0)= M_0$. To solve this equation it is necessary to know $\mu(I)$, $P(t)=P(M(t))$ and $\alpha$. In the rest of this paper, instead of directly solving Eq. (\ref{diff}), we estimate $\mu(I)$ and then extract an expression for $P(M(t))$ based on the condition that the flow rate must be constant throughout the discharge, as observed experimentally. This allows to fit the value of $\alpha$ using pressure data from DEM simulations of the discharge. 

To estimate $\mu(I)$, consider that if $R_s \rightarrow \infty$, then $I\rightarrow 0$. Da Cruz  \textit{et al}. have shown that in the quasistatic limit (i.e., $I\rightarrow 0$) $\mu(I=0)\approx 0.26$ for all material properties if the grain--grain friction coefficient is above 0.4 \cite{dacruz2005}. We have carried out DEM simulations using YADE \cite{yade}, for a silo where $R_s=15$ mm and $D_o=15$ mm, and a granular sample with $d=1$ mm, $\rho=2000$ kg/m$^3$,  $\rho_b=1160$ kg/m$^3$ and the friction coefficient of the grain--grain and grain--wall interaction is $0.55$. The inertial number estimated as $I=\frac{v}{R_s}\frac{d}{\sqrt{P/\rho}}$  varies during the discharge in the range $9.8\times 10^{-3} <I< 1.6 \times 10^{-2}$, which is in the quasistatic limit consistent with an effective friction coefficient $0.26$ \cite{dacruz2005}. 

\begin{figure}
 \begin{center}
  \includegraphics[width=0.99\columnwidth]{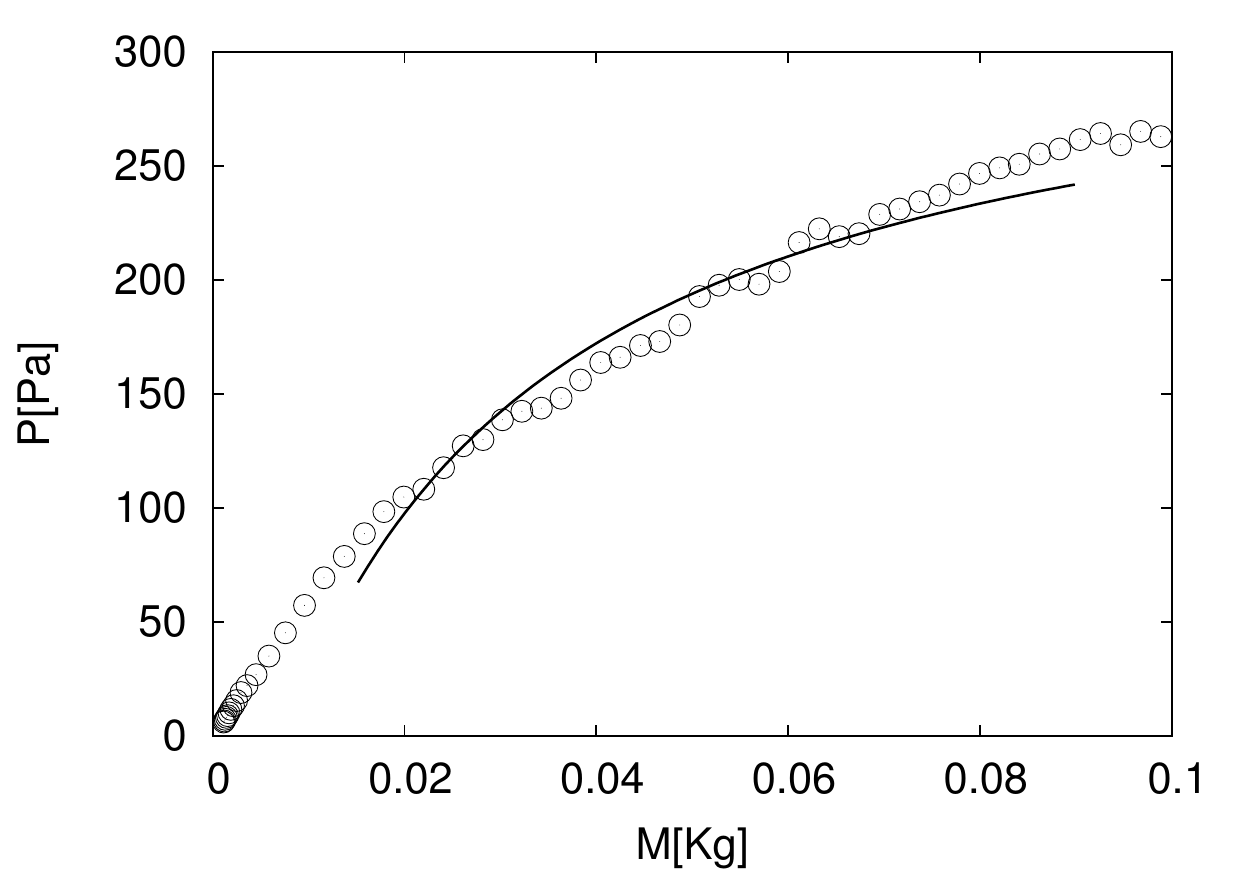}
 \end{center}
\caption{Pressure (isotropic component $P=\text{Tr}(\sigma)/3$ of the stress tensor) as a function of the mass $M$ inside the silo during the discharge. Results obtained via a DEM simulation for a silo (symbols) and fit using Eq. (\ref{pressure}) (line).}
\label{fig-pressure}
\end{figure}

As we mentioned, Eq. (\ref{diff}) does not show the classical $5/2$ exponent for $D_o$, nor the correction $-kd$ for the ``empty annulus''. However, the expression under the square root in Eq. (\ref{diff}) must compensate the difference between $D_o^2$ and the traditional Beverloo factor $(D_o-kd)^{5/2}$. During the discharge it is known that $Q$ is constant. Hence, the radicand in  Eq. (\ref{diff}) must be also a constant $\gamma$. Therefore, the pressure can be written as

\begin{eqnarray}  \label{pressure}
P(t) = \left[\frac{M(t)}{\rho_b A_s}-\gamma \right] \frac{g \rho^2_b A^2_s}{\mu(I)[2\pi R_sM(t)+\alpha\rho_bA^2_s]}.
\end{eqnarray}

If the radicand $\gamma$ has to compensate the difference with the Beverloo rule, then $\gamma = (D_o-kd)^{5}/D_o^4$. For our simulations with $D_o=15$ mm and $d=1$ mm, and considering a typical value of $k=1.4$, we obtain $\gamma \approx 9.2\times 10^{-3}$. 

Figure \ref{fig-pressure} shows a fitting of Eq. (\ref{pressure}) to the aforementioned DEM data of the silo discharge. The pressure is calculated as the average of $\text{Tr}(\sigma)/3$ over the entire volume of the granular column a any given time during the discharge. Here, $\sigma$ is the stress tensor. We took, following the discussion above, $\mu(I)=0.26$ and $\gamma=9.2\times 10^{-3}$. As we can see, the only fitting parameter is $\alpha$ whose best fit value is $\alpha=3.6 \pm 0.2$. The fit is only applied to a section of the process since the assumption of a constant flow rate is not justified in the final stage of the discharge. The fit is qualitatively fair. In particular, Eq. (\ref{diff}) has a saturation of the pressure that is much slower than the traditional Janssen law for static silos. This is consistent with the results found here for the simulations and also results reported in experiments of discharging silos \cite{perge2012}.  

We speculate that the value of $\alpha$ found may be ``universal'' in the sense that it does not depend significantly on $D_o$, $R_s$, and material properties. However, this remains a subject of investigation. 

\section{Conclusions}

We have shown that an energy balance, coupled with a constitutive equation for the rheology of a granular sample leads to a simple differential equation for the flow rate of a discharging silo. The equation resembles the Beverloo rule and yields a proportionality constant $C = \frac{\pi\sqrt{2}}{8} \approx 0.56$ which is very close to the experimental fitted value of $0.58$. 

The assumption of a constant flow rate leads to an expression for the pressure in the silo that is more suitable for these dynamic conditions than the Janssen law derived for static silos. We have to bear in mind, however, that the pressure in the silo is not homogeneous and a more elaborated analysis should include the local pressure in the description rather than the global pressure as we have done in this work.

The derived differential equation allows to explain the known phenomenology of silo discharge without the need of heuristic postulates such as the ``free fall'' arch and the ``empty annulus''. Moreover, this equation also opens the way to consider more complex situations in silo discharge, such as forced flow or vibrated discharges, that have not been explored from a theoretical perspective.

\section*{Acknowledgements}
This work has been sup-ported by ANPCyT (Argentina) through grant PICT 2012-2155 and UTN (Argentina) through grants PID MAUTNLP0002184 and PID IFI1871.

%
%
%

\end{document}